\documentclass[useAMS,usenatbib]{pasj01}
\usepackage{times,aas_macros}
\usepackage{graphicx}
\usepackage{rotating}
\input{epsf}

\begin{document}

\title{Does AGN Fraction Depend on Redshift or Luminosity? An Extinction-Free Test by 18-band Mid-infrared SED Fitting in the AKARI NEP Wide Field}
\author{Chia-Ying \textsc{Chiang}\altaffilmark{1,2}}
\altaffiltext{1}{Institute of Astronomy, National Tsing Hua University, No. 101, Section 2, Kuang-Fu Road, Hsinchu City 30013, Taiwan}
\altaffiltext{2}{Department of Physics \& Astronomy, Wayne State University, 666 W. Hancock, Detroit, MI 48202, USA}
\altaffiltext{3}{Institute of Space and Astronautical Science, Japan Aerospace Exploration Agency, Yoshinodai 3-1-1, Sagamihara, Kanagawa 229-8510, Japan}
\altaffiltext{4}{Tokyo University of Science, 1-3 Kagurazaka, Shinjuku-ku, Tokyo 162-8601, Japan}
\email{ft8320@wayne.edu}
\Received{15-Jun-2018} \Accepted{28-Jan-2019}

\author{Tomotsugu \textsc{Goto}\altaffilmark{1}}
\author{Tetsuya \textsc{Hashimoto}\altaffilmark{1}}
\author{Seong-Jin \textsc{Kim}\altaffilmark{1}}
\author{Hideo \textsc{Matsuhara}\altaffilmark{3}}
\author{Nagisa \textsc{Oi}\altaffilmark{4}}

\maketitle

\begin{abstract}
Revealing what fraction of galaxies harbor AGN is central in understanding black hole accretion 
history of the Universe. However, optical and soft X-ray surveys miss the most highly obscured AGNs. 
Infrared (IR), instead, is more robust against absorption. Previous IR photometric surveys, 
however, only had 4 or 5 filters in mid-IR. Our AKARI North Ecliptic Pole (NEP) wide field 
sample has 18 filters in mid-IR (9 from AKARI, 4 from WISE, and 5 from Spitzer), for the 
first time, allowing a sophisticated mid-IR SED fitting diagnosis for a statistical number of 
sources (89178 over 5.4 deg$^2$). By using a SED fitting technique, we investigate the 
evolution of AGN fraction as a function of redshift and IR (8-1000 $\mu$m) luminosity in an extinction-free way. 
We found that the AGN fraction (F$_{\rm AGN}$) shows no sign of strong redshift evolution. 
Instead,  F$_{\rm AGN}$ increases with increasing IR luminosity in all redshifts bins ($0<z<2$).
\end{abstract}

\section{Introduction}
Active Galactic Nuclei (AGNs) are powered by central accreting supermassive black holes, 
and cast impact on the environment via AGN feedback processes \citep{Kauffmann00,DiMatteo05,Fabian12}. It is essential to understand
AGNs in order to probe cosmological evolution of supermassive black holes and the 
universe. In addition, AGN fraction is important when studying luminosity functions. Researchers
have been investigating various surveys to study AGNs for decades.
Although AGNs can be bright in a wide range 
of wavelengths, most surveys are severely biased towards unobscured (Type 1) AGNs.
A large population of obscured (Type 2) AGNs are missed in optical and soft X-ray surveys due to 
obscuration caused by dust and gas. In this case, radio, hard X-ray, and mid-infrared (IR)
selections are promising for detecting Type 2 AGNs. However, only $\sim$10\% of AGNs
are radio-loud, and current hard X-ray satellites (i.e. \emph{Swift} BAT, \emph{NuSTAR}, etc.) have limited sensitivity. It is difficult for radio 
and hard X-ray surveys to reach a large sample size to overcome cosmic variance. As warm dust emission 
dominates the mid-IR band (2-24 $\mu$m), and mid-IR satellites provide sensitive full-sky surveys,
mid-IR selection is an effective, nearly free of extinction way
to identify AGNs. In past few years there has been numerous 
studies using mid-IR colors to find AGNs in the local universe and at high-redshift \citep{Lacy04,Lacy07,Stern12}.

As star-forming galaxies (SFGs) are bright in mid-IR bands as well, it has been of 
great challenge to distinguish SFGs and AGNs. In
mid-IR, a SFG has polycyclic aromatic hydrocarbon (PAH)
broad emission line features at 3.3, 6.2, 7.7, 8.6, 11.2, and 12.7 $\mu$m \citep{Puget85,Allamandola89}, 
while an AGN presents a red power-law ($f_{\nu}\propto\nu^{-\alpha}$) spectrum 
with typical $\alpha\leq-0.5$ \citep{AH06}. 
\citet{Laurent00} showed that the mid-IR Spectral Energy Distributions (SEDs) of
SFGs and AGNs are distinct enough for separation, and some following work 
also confirmed that mid-IR photometry provides robust SFG-AGN diagnostics
\citep{Stern05,Jarrett11,Magdis13}. Survey data obtained by \emph{Spitzer} infrared 
telescope and Wide field Infrared Survey Explorer (\emph{WISE}) lead to an explosion
of research to search and study AGNs \citep{Lacy04,AH06,Lacy07,Lacy13,Lacy15,Eckart10,
Stern05,Donley12,Mateos12,Stern12,Assef13}. 
However, both \emph{WISE} and \emph{Spitzer} have limited available filters and 
gaps (between 12-22 $\mu$m in \emph{WISE} and 8-24 $\mu$m in \emph{Spitzer}) 
in the mid-IR wavelength range, which make SFG-AGN diagnostics difficult for sources 
at certain redshift ranges. PAH features of high redshift sources may fall into the 
wavelength gaps and make AGN selection difficult. For instance, the 8 to 24 $\mu$m
gap is important for the diagnosis of AGN and SFG at $0.5<z<1.5$. The infrared satellite 
\emph{AKARI} \citep{Murakami07} has 9-band mid-IR filters and does not
suffer from the lack of continuous mid-IR data. 
The \emph{AKARI} North Ecliptic Pole (NEP) survey covers a field of 5.4 deg$^{2}$
and will remain the only survey with continuous mid-IR data until
the James Webb Space Telescope (\emph{JWST}) performs a similar one.

\citet{Huang17} (hereafter H17) used the \emph{AKARI} NEP deep survey \citep{Matsuhara06} covering a field of 0.57 
deg${^2}$ catalogue (containing $\sim$5800 sources) to select AGNs using the SED fitting 
technique. By combining \emph{AKARI}, \emph{WISE} and \emph{Spitzer} 
data with 18 bands in total and fitting 25 empirical models to the data, the 
authors recovered more X-ray detected AGNs than previous work by 
$\sim$20\%. Previous research using \emph{WISE} or \emph{Spitzer}
data performed SED fitting with limited models \citep{Assef08,Assef10,Assef13,Chung14}. 
Thanks to the continuous coverage of mid-IR band by \emph{AKARI}, 
more detailed models can be used to help AGN selection at high redshift.
In this work, we use the \emph{AKARI} NEP wide field (covering 5.4 deg$^{2}$)
catalogue for AGN selection. We will provide detailed information of 
the catalogue and data analysis in section \ref{section_da}, and present 
results in section \ref{results}. We will discuss the results compared with
other work in section \ref{discussion}, and summarize the paper in section
\ref{conclusion}.

\section{Data Analysis} \label{section_da}

\subsection{Data}
We used the catalogue from the NEP wide field survey \citep{Kim12} of 
\emph{AKARI} in this work. The survey was carried out
on a circular area of $\sim$5.4 deg$^2$ and is part of the large area survey 
programs of \emph{AKARI}, which provides crucial FIR bands to measure SED accurately \citep{Kim15}. 
The \emph{AKARI} infrared camera (IRC; \cite{Onaka07}) offers coverage from near to mid-infrared 
(2-24 $\mu$m) wavelength bands. There are 9 filters (N2, N3, N4, S7, S9W, S11, L15,
L18W, and L24) corresponding to 2.4, 3.2, 
4.1, 7, 9, 11, 15, 18 and 24 $\mu$m effective wavelength. The instrument detection limits 
are 11, 9, 10, 30, 34, 57, 87, 93, and 256 $\mu$Jy in the above 9 bands, respectively. 
The detection limits of the NEP survey are 20.9, 21.1, 21.1, 19.5, 19.4, 19.0, 18.6, 18.7 and 17.8 
in AB magnitude \citep{Kim12}.
In this catalogue data of 4 mid-IR bands from \emph{WISE} (3.4, 4.6, 12 and 22 $\mu$m)
and 5 mid-IR bands from \emph{Spitzer} (3.6, 4.5, 5.8, 8.0 and 24 $\mu$m) are included,
forming a sample with 18 mid-IR bands in total.

\begin{figure}
\begin{center}
\leavevmode \epsfxsize=8.5cm \epsfbox{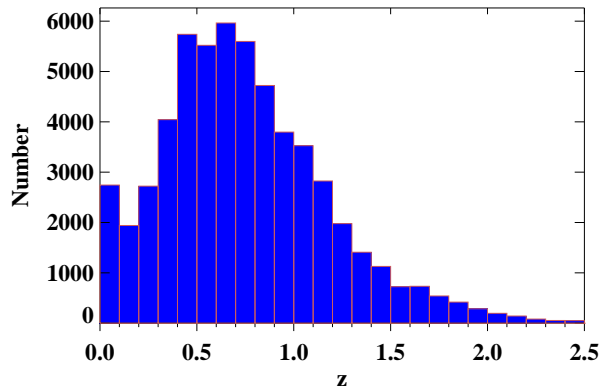}
\end{center}
\caption{Photometric redshift distribution of the catalogue with bin size of 0.1. Stars are excluded from the figure.
}
\label{distribution}
\end{figure}

The \emph{AKARI} NEP survey detected a large number of IR sources, yet
H17 only used sources in the deep field, which is a small part of it due to 
limited optical/UV coverage by the Galaxy Evolution Explorer (\emph{GALEX}) 
and the Canada France Hawaii Telescope (\emph{CFHT}). An
optical survey covering the \emph{AKARI} NEP wide field using \emph{Subaru's}
Hyper Suprime-Cam (HSC; \cite{Miyazaki12}) has been carried out recently.
The HSC coverage \citep{Goto17} increases the number of sources with optical data to 89178 
in the catalogue, which is the largest sample to date. Photometric redshifts can 
be measured accurately from the data of the five HSC optical bands 
($g$, $r$, $i$, $z$ and $y$; Oi et al., submitted to this volume). 
Fig. \ref{distribution} shows the redshift distribution 
of the sample. In this work we only focus on galaxies, and after excluding stars in the 
catalogue, there are 57106 galaxies remaining in the sample.

\subsection{SED Fitting}
We performed SED fitting using the \textit{Le PHARE}
\footnote{http://www.cfht.hawaii.edu/$\sim$arnouts/LEPHARE/lephare.html} \citep{Arnouts99,Ilbert06} code.
The \citet{Lagache03} template was used to measure infrared luminosities in the catalogue.
As galaxy types were not given by the Lagache template, we used the \citet{Polletta07} 
templates, which contains 3 elliptical (type number 1-3), 7 spiral (type number 4-10), 
6 starburst (type number 11-16), 7 AGN (type number 17-22 and 25), and 2 AGN-
Starburst composite (type number 23-24) models and is part of the Spitzer Wide-area Infrared Extragalatic 
(SWIRE) templates\footnote{http://www.iasf-milano.inaf.it/$\sim$polletta/templates/swire\_templates.html}, 
to determine the spectral type of galaxies. Each type number represents a type of galaxy in the templates,
and numbers 17-25 are of AGN SEDs (see also Table 1 in H17 for details). 
As photometric redshifts have been measured in the sample (Oi et al., submitted to this volume), 
we fixed redshifts with the ZFIX=YES setting when 
performing SED fitting using the \textit{Le PHARE} code. 
The Calzetti extinction law \citep{Calzetti00} has been used in the fitting.
As a result, \emph{Le PHARE} found best-fitting models for 16464 
sources in the sample via SED fitting. In the following we focus on
the 16464 sources and hereafter refer it as ``the SED sample".

\section{Results} \label{results}

\begin{figure}
\begin{center}
\leavevmode \epsfxsize=8.5cm \epsfbox{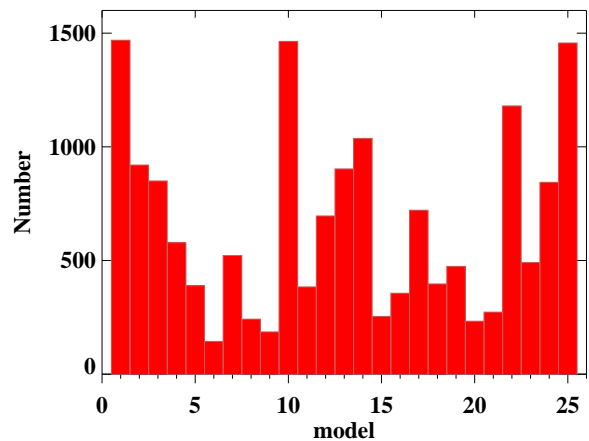}
\end{center}
\caption{Distribution of best-fitting models in the SED sample. Numbers 1-3 represent elliptical galaxy models;
4-10 spiral galaxy models; 11-16 starburst galaxy models; 17-25 AGN models.
}
\label{type}
\end{figure}

We plot the distribution of best-fitting models in Fig. \ref{type}. 
There are 6070 AGNs selected among the total 16464 galaxies,
giving an overall AGN fraction of $36.9\pm0.5$\%, which is mildly
higher than the number of $29.6\pm0.8$\% reported in H17 using 
the NEP deep field catalogue. Our sample size is $\sim$3.5 times
larger than that used in H17. Surprisingly, a large number (3239 in total) of elliptical
galaxies (type number 1-3) are detected as this class of galaxies are
usually faint in the mid-IR. Type numbers 17-19 and 24-25 belong
to Type 2 AGNs, and it seems mid-IR observations are good at 
selecting obscured AGNs as expected.

\subsection{AGN Fraction as a Function of Redshift} 

\begin{figure}
\begin{center}
\leavevmode \epsfxsize=8.5cm \epsfbox{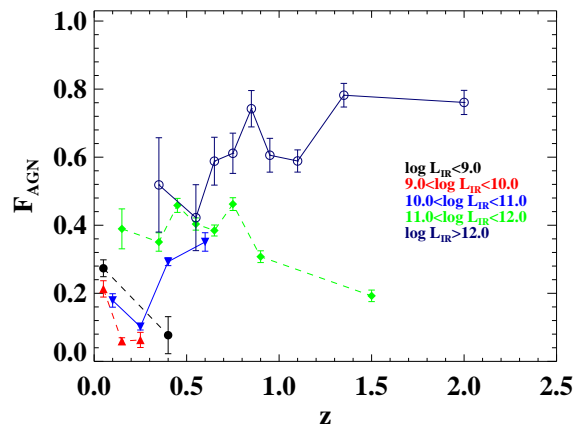}
\end{center}
\caption{AGN fraction as a function of photometric redshift. Black solid dots represent the
log L$_{\rm IR}<9.0$ luminosity bin; red triangles stand the $9.0<$ log $L_{\rm IR}<10.0$ bin; blue triangles are of the $10.0<$ log $L_{\rm IR}<11.0$ bin;
green diamonds represent the $11.0<$ log $L_{\rm IR}<12.0$ bin; dark blue circle belong to the log L$_{\rm IR}>12.0$ bin. 
Errors in this and following figures are calculated by Poisson distribution. It seems that the result does
not reveal significant redshift dependence.
}
\label{FA_z}
\end{figure}

We separate the SED sample into a few luminosity bins
and plot the evolution of AGN fraction as a function of photometric redshift (see Fig. \ref{FA_z}).
It is clearly seen that all luminosity bins show a relatively constant or decreasing 
trend at $z<0.5$ except the $10<$ log $L_{\rm IR}<11$ bin. In the regime
of $z>0.5$, the $10<$ log $L_{\rm IR}<11$ and log $L_{\rm IR}>12$ bins
show an increasing trend, while the $11<$ log $L_{\rm IR}<12$ bin displays
a decreasing trend. Different luminosity bins show different behaviors, and
from our results it seems that there may be no strong redshift dependence in
AGN fraction evolution. 

\subsection{AGN Fraction as a Function of Luminosity} 

\begin{figure}
\begin{center}
\leavevmode \epsfxsize=8.5cm \epsfbox{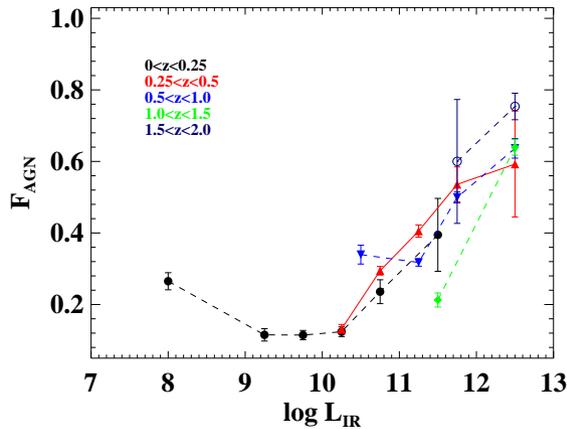}
\end{center}
\caption{AGN fraction as a function of luminosity. Black solid dots represent the
$0<z<0.25$ redshift bin; red triangles stand the $0.25<z<0.5$ bin; blue triangles are of the $0.5<z<1.0$ bin;
green diamonds represent the $1.0<z<1.5$ bin; dark blue circle belong to the $1.5<z<2.0$ bin.
All bins show similar behaviors, implying that
there may be more luminosity dependence than redshift dependence.
}
\label{FA_L}
\end{figure}

In order to check if AGN fraction has luminosity dependence, we
divide the SED sample into a few groups of different redshift range, 
and plot the evolution of AGN fraction as a function of luminosity in
Fig. \ref{FA_L}. We find that all redshift bins show an increasing trend
at log $L_{\rm IR}>10$. Furthermore, all bins seem to reach similar
values of AGN fraction, which again reveals that there is no significant
redshift dependence. Instead, the result implies that there is more 
luminosity dependence than redshift dependence in AGN fraction.
Note that most luminosity bins reach a fraction of $\sim$0.6 at high luminosities.

\section{Discussion} \label{discussion}

\subsection{AGN Fraction}
We show that there is possible luminosity dependence in the 
evolution of AGN fraction, which is consistent with the result 
reported in previous literature.
\citet{Goto05} investigated 4828 Infrared Astronomical Satellite (\emph{IRAS})
galaxies and selected AGNs using the optical emission-line ratios, and
found that the AGN fraction increases as a function of $L_{\rm IR}$.
Similar trends have been reported by \citet{Yuan10}, \citet{Kartaltepe10},
 \citet{Goto11}, and \citet{Weigel18} using various local samples.
These results imply that luminous infrared galaxies (LIRGs; $11<$ log $L_{\rm IR}<12$) and ultraluminous
infrared galaxies (ULIRGs; $12<$ log $L_{\rm IR}<13$) are more likely to be powered by
AGNs. AGN fractions reported in previous literature
also reach $\sim$0.6, which is consistent with the result of the present work. Note that 
previous research only used samples in the local ($z<0.3$) universe, and our
work extends the study to trace behaviors of galaxies at higher redshifts and compares
results between different branches of redshifts. We show that the luminosity dependence
of AGN fraction may be global.

\begin{figure}
\begin{center}
\leavevmode \epsfxsize=8.5cm \epsfbox{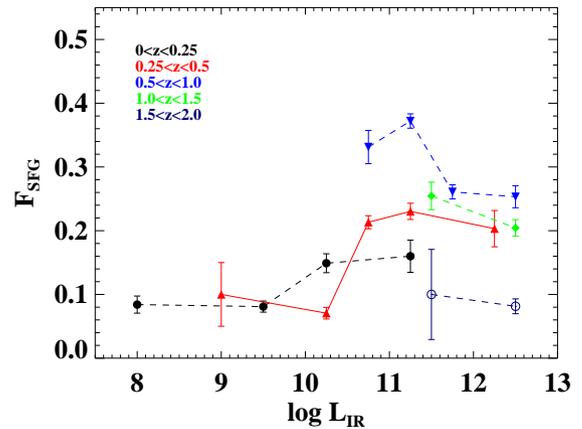}
\leavevmode \epsfxsize=8.5cm \epsfbox{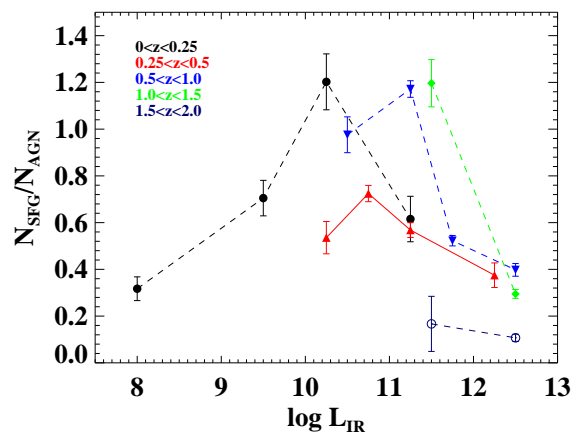}
\end{center}
\caption{Top: SFG fraction as a function of luminosity. Black solid dots represent the
$0<z<0.25$ redshift bin; red triangles stand the $0.25<z<0.5$ bin; blue triangles are of the $0.5<z<1.0$ bin;
green diamonds represent the $1.0<z<1.5$ bin; dark blue circle belong to the $1.5<z<2.0$ bin. Note that all bins show mild decreasing trends at high luminosities.
Bottom: N$_{\rm SFG}$/N$_{\rm AGN}$ ratio as a function of luminosity. It can be seen that AGN dominates at 
high luminosities.
}
\label{FS_L}
\end{figure}

The reason that the AGN fraction is high among LIRGs can be explained by
one of the merger scenarios for LIRGs \citep{Sanders88,Dasyra06}. When two galaxies merge, they lose orbital
energy and angular momentum to tidal features and/or shocks, which trigger star
formation \citep{Kennicutt87,Liu95,Barnes04}. These activities heat surrounding dust and produce strong far-infrared (FIR)
radiation. It then reaches an ultralumious IR stage which is powered by starburst 
or AGNs. The merger finally evolves into an optically bright QSO when star formation
activities decline. In this scenario AGNs dominate luminous IR sources during
final merger stages. We plot the fraction of SFGs as a function of luminosity in the top panel of Fig. \ref{FS_L}.
It can be seen that fractions of SFGs of all redshift bins show mild decreasing
trends at high $L_{\rm IR}$. The lower panel of Fig. \ref{FS_L} shows the 
value of N$_{\rm SFG}$/N$_{\rm AGN}$, which is the number of SFGs over the number of AGNs, and the clear decreasing trends at high luminosities further
supports the prediction of the theory.

\subsection{Other SED Fitting Templates}

H17 performed SED fitting of the NEP deep field sample
using the templates provided by \citet{Brown14} as well, and found that
the results are consistent with those given by the Polletta's template set.
The Brown's template set contains 100 models in total, and provides a set 
of 37 SFG, 11 AGN and 22 AGN-Starburst composite models which allow more detailed classifications.
The \emph{LePHARE} code determined types for 16417 sources in the 
catalogue, which is comparable to the number obtained using the Polletta's template set and all of them are included in the SED sample. 
There are 5427 SFGs, 2083 AGNs and 3890 AGN-Starburst composites classified by
the Brown's template set. Note that 3097 galaxies are classified as elliptical galaxies, which
is comparable with the large number of elliptical galaxies classified by the Polletta's template set.
Around 77 \% of Polletta elliptical galaxies are also classified as elliptical galaxies by the Brown's template set.
Only 33 \% of SFGs, AGNs and AGN-Starburst composites have the same classification in both Pollettas's and Brown's template sets.
However, the overall AGN fraction (including AGNs and composites) is $36.4\pm0.5$\%, which
is remarkably similar to the number obtained using the Polletta's template set. It is not
surprising that the Brown's template set is capable of more detailed classification because there are
a lot more models than those of the Polletta's template set. Note that the Brown's template set
does not determine types for more galaxies than the Polletta's template set, indicating that both template sets
perform similarly on the \emph{AKARI} survey data. Given that not all sources in the SED sample are detected
by every of the 18 mid-IR band, increasing models of SED fitting does not necessarily lead to more information.
In this sense, results obtained using different templates should not differ
significantly and the Polletta's template set remains reliable in our sample.
Future development is expected when \emph{JWST} data are available.

\begin{figure}
\begin{center}
\leavevmode \epsfxsize=8.5cm \epsfbox{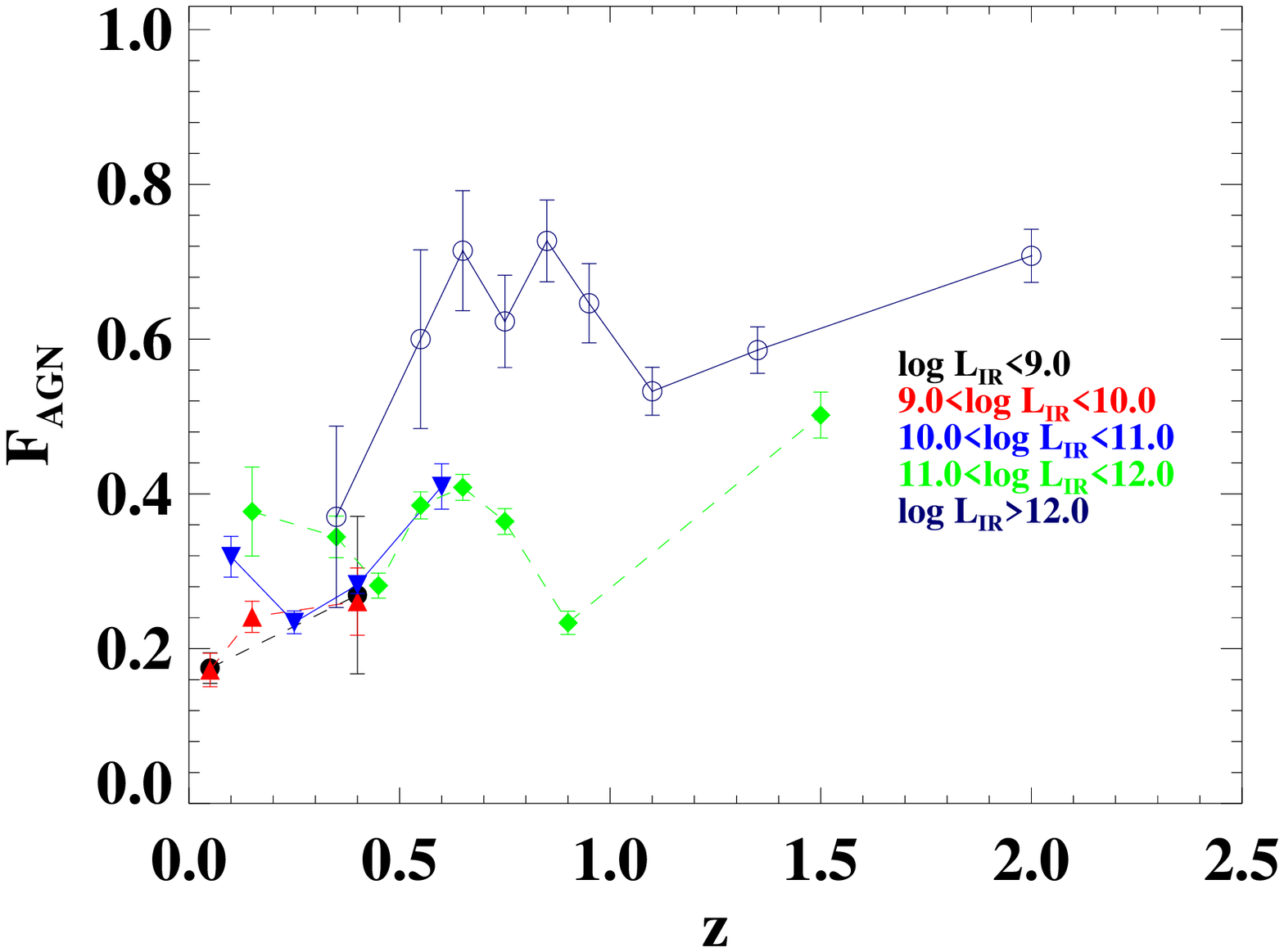}
\leavevmode \epsfxsize=8.5cm \epsfbox{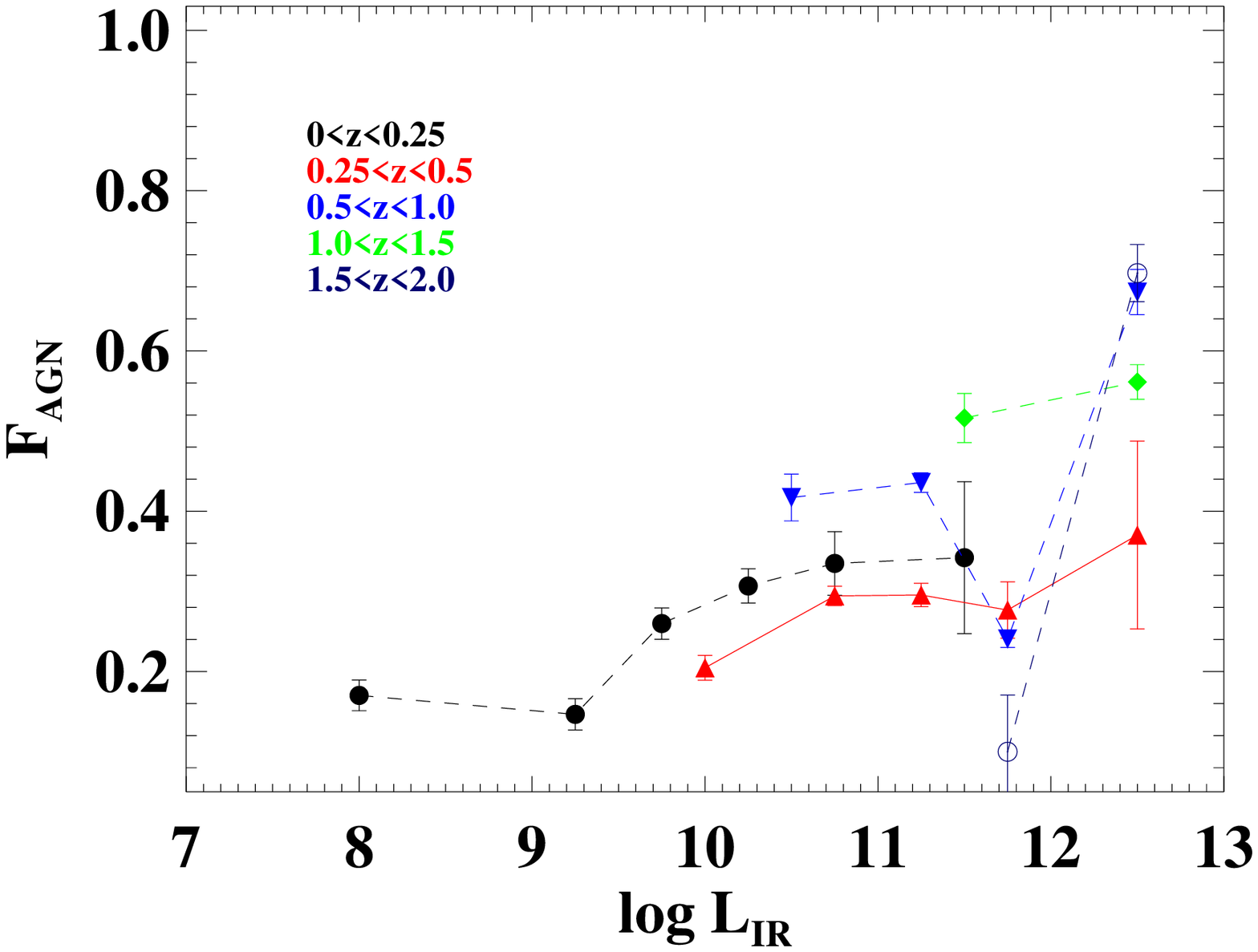}
\leavevmode \epsfxsize=8.5cm \epsfbox{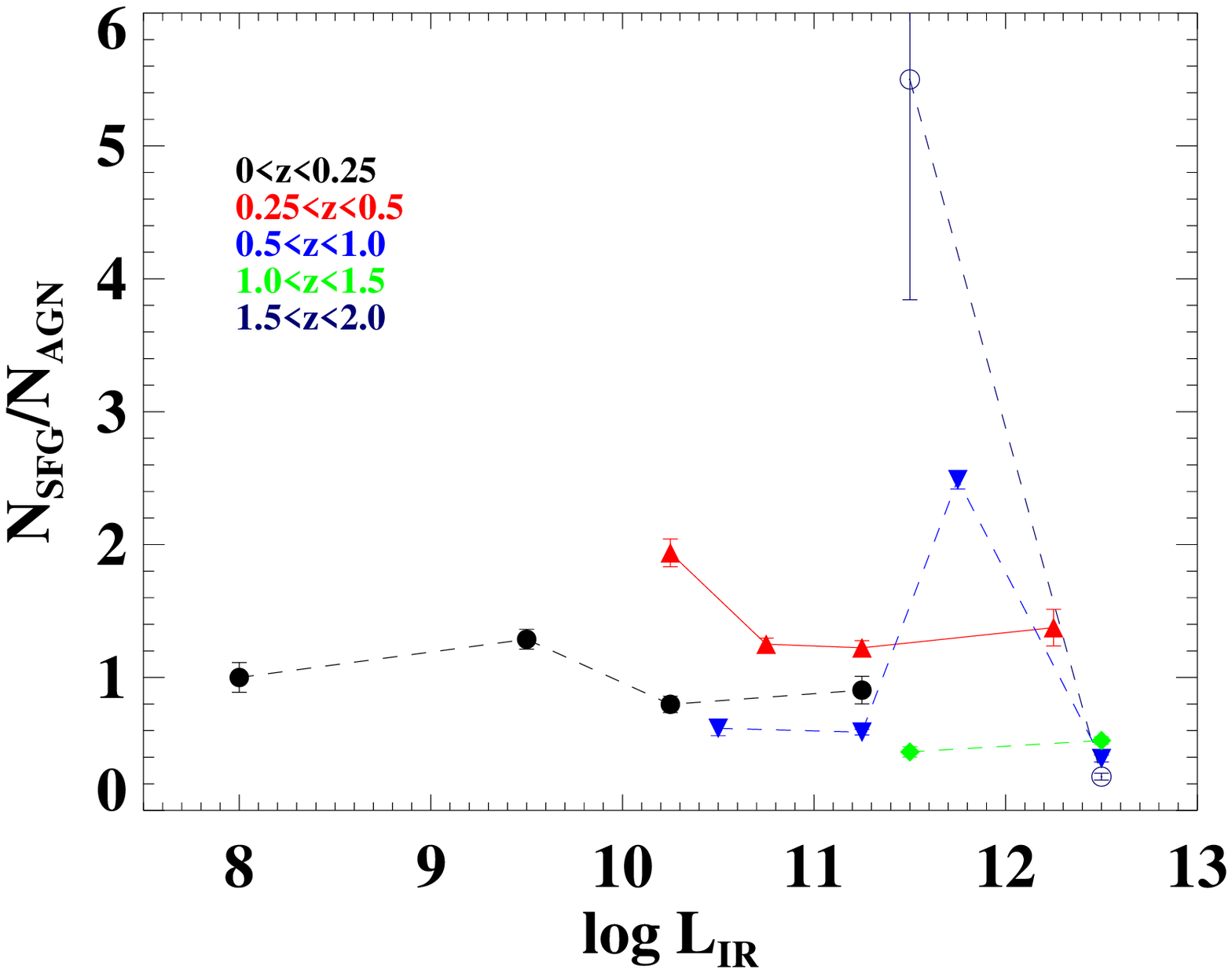}
\end{center}
\caption{Results obtained using the Brown template.
Top: AGN fraction as a function of photometric redshift; middle: AGN fraction
as a function of luminosity; bottom: N$_{\rm SFG}$/N$_{\rm AGN}$ ratio as a function of luminosity. Symbols
are the same as those used in Fig. \ref{FA_z}, Fig. \ref{FA_L}, and Fig. \ref{FS_L}, respectively.
}
\label{FA_Brown}
\end{figure}

Fig. \ref{FA_Brown} shows results obtained using the Brown's template set. It can be seen that
the AGN fraction still shows no significant redshift dependence. The luminosity
dependence is not as strong as that showed in the Polletta results, but most bins do
show significant increase at high luminosities.

\section{Conclusions} \label{conclusion}

We used the unique \emph{AKARI} NEP wide field survey sample with
18 mid-IR bands of data to perform SED fitting and select AGNs.
Our results indicate that AGN fraction seems to show
stronger luminosity dependence than redshift dependence.
This is consistent with results reported in previous literature
and the theoretical model of ULIRGs. We also examined the
fractions of SFGs and found mild decreasing trends at high
luminosities, which is in agreement with theoretical predictions.
The major improvement from previous research is that our sample
contains both local and high-redshift galaxies, the other work 
only used local samples. Although further improvement is expected
with the coming launch of \emph{JWST}, \emph{AKARI} NEP survey
will remain the best sample to study luminosity function, AGN fractions,
etc., in decades.

\section*{Acknowledgements}
We acknowledge the support by the Ministry of Science and Technology (MoST) of Taiwan through grant105-2112-M-007-003-MY3.
CYC thanks the support by the MoST Leaders in Future Trends (LiFT) program.

\bibliographystyle{apj}
\bibliography{NEP}

\begin{thebibliography}{}
\expandafter\ifx\csname natexlab\endcsname\relax\def\natexlab#1{#1}\fi

\bibitem[{{Allamandola} {et~al.}(1989){Allamandola}, {Tielens}, \&
  {Barker}}]{Allamandola89}
{Allamandola}, L.~J., {Tielens}, A.~G.~G.~M., \& {Barker}, J.~R. 1989, \apjs,
  71, 733

\bibitem[{{Alonso-Herrero} {et~al.}(2006){Alonso-Herrero},
  {P{\'e}rez-Gonz{\'a}lez}, {Alexander}, {Rieke}, {Rigopoulou}, {Le Floc'h},
  {Barmby}, {Papovich}, {Rigby}, {Bauer}, {Brandt}, {Egami}, {Willner}, {Dole},
  \& {Huang}}]{AH06}
{Alonso-Herrero}, A., {P{\'e}rez-Gonz{\'a}lez}, P.~G., {Alexander}, D.~M.,
  {et~al.} 2006, \apj, 640, 167

\bibitem[{{Arnouts} {et~al.}(1999){Arnouts}, {Cristiani}, {Moscardini},
  {Matarrese}, {Lucchin}, {Fontana}, \& {Giallongo}}]{Arnouts99}
{Arnouts}, S., {Cristiani}, S., {Moscardini}, L., {et~al.} 1999, \mnras, 310,
  540

\bibitem[{{Assef} {et~al.}(2008){Assef}, {Kochanek}, {Brodwin}, {Brown},
  {Caldwell}, {Cool}, {Eisenhardt}, {Eisenstein}, {Gonzalez}, {Jannuzi},
  {Jones}, {McKenzie}, {Murray}, \& {Stern}}]{Assef08}
{Assef}, R.~J., {Kochanek}, C.~S., {Brodwin}, M., {et~al.} 2008, \apj, 676, 286

\bibitem[{{Assef} {et~al.}(2010){Assef}, {Kochanek}, {Brodwin}, {Cool},
  {Forman}, {Gonzalez}, {Hickox}, {Jones}, {Le Floc'h}, {Moustakas}, {Murray},
  \& {Stern}}]{Assef10}
---. 2010, \apj, 713, 970

\bibitem[{{Assef} {et~al.}(2013){Assef}, {Stern}, {Kochanek}, {Blain},
  {Brodwin}, {Brown}, {Donoso}, {Eisenhardt}, {Jannuzi}, {Jarrett}, {Stanford},
  {Tsai}, {Wu}, \& {Yan}}]{Assef13}
{Assef}, R.~J., {Stern}, D., {Kochanek}, C.~S., {et~al.} 2013, \apj, 772, 26

\bibitem[{{Barnes}(2004)}]{Barnes04}
{Barnes}, J.~E. 2004, \mnras, 350, 798

\bibitem[{{Brown} {et~al.}(2014){Brown}, {Moustakas}, {Smith}, {da Cunha},
  {Jarrett}, {Imanishi}, {Armus}, {Brandl}, \& {Peek}}]{Brown14}
{Brown}, M.~J.~I., {Moustakas}, J., {Smith}, J.-D.~T., {et~al.} 2014, \apjs,
  212, 18

\bibitem[{{Calzetti} {et~al.}(2000){Calzetti}, {Armus}, {Bohlin}, {Kinney},
  {Koornneef}, \& {Storchi-Bergmann}}]{Calzetti00}
{Calzetti}, D., {Armus}, L., {Bohlin}, R.~C., {et~al.} 2000, \apj, 533, 682

\bibitem[{{Chung} {et~al.}(2014){Chung}, {Kochanek}, {Assef}, {Brown}, {Stern},
  {Jannuzi}, {Gonzalez}, {Hickox}, \& {Moustakas}}]{Chung14}
{Chung}, S.~M., {Kochanek}, C.~S., {Assef}, R., {et~al.} 2014, \apj, 790, 54

\bibitem[{{Dasyra} {et~al.}(2006){Dasyra}, {Tacconi}, {Davies}, {Naab},
  {Genzel}, {Lutz}, {Sturm}, {Baker}, {Veilleux}, {Sanders}, \&
  {Burkert}}]{Dasyra06}
{Dasyra}, K.~M., {Tacconi}, L.~J., {Davies}, R.~I., {et~al.} 2006, \apj, 651,
  835

\bibitem[{{Di Matteo} {et~al.}(2005){Di Matteo}, {Springel}, \&
  {Hernquist}}]{DiMatteo05}
{Di Matteo}, T., {Springel}, V., \& {Hernquist}, L. 2005, \nat, 433, 604

\bibitem[{{Donley} {et~al.}(2012){Donley}, {Koekemoer}, {Brusa}, {Capak},
  {Cardamone}, {Civano}, {Ilbert}, {Impey}, {Kartaltepe}, {Miyaji}, {Salvato},
  {Sanders}, {Trump}, \& {Zamorani}}]{Donley12}
{Donley}, J.~L., {Koekemoer}, A.~M., {Brusa}, M., {et~al.} 2012, \apj, 748, 142

\bibitem[{{Eckart} {et~al.}(2010){Eckart}, {McGreer}, {Stern}, {Harrison}, \&
  {Helfand}}]{Eckart10}
{Eckart}, M.~E., {McGreer}, I.~D., {Stern}, D., {Harrison}, F.~A., \&
  {Helfand}, D.~J. 2010, \apj, 708, 584

\bibitem[{{Fabian}(2012)}]{Fabian12}
{Fabian}, A.~C. 2012, \araa, 50, 455

\bibitem[{{Goto}(2005)}]{Goto05}
{Goto}, T. 2005, \mnras, 360, 322

\bibitem[{{Goto} {et~al.}(2011){Goto}, {Arnouts}, {Malkan}, {Takagi}, {Inami},
  {Pearson}, {Wada}, {Matsuhara}, {Yamauchi}, {Takeuchi}, {Nakagawa}, {Oyabu},
  {Ishihara}, {Sanders}, {Le Floc'h}, {Lee}, {Jeong}, {Serjeant}, \&
  {Sedgwick}}]{Goto11}
{Goto}, T., {Arnouts}, S., {Malkan}, M., {et~al.} 2011, \mnras, 414, 1903

\bibitem[{{Goto} {et~al.}(2017){Goto}, {Toba}, {Utsumi}, {Oi}, {Takagi},
  {Malkan}, {Ohayma}, {Murata}, {Price}, {Karouzos}, {Matsuhara}, {Nakagawa},
  {Wada}, {Serjeant}, {Burgarella}, {Buat}, {Takada}, {Miyazaki}, {Oguri},
  {Miyaji}, {Oyabu}, {White}, {Takeuchi}, {Inami}, {Perason}, {Malek},
  {Marchetti}, {Lee}, {Im}, {Kim}, {Koptelova}, {Chao}, {Wu}, {AKARI NEP Survey
  Team}, \& {AKARI All Sky Survey Team}}]{Goto17}
{Goto}, T., {Toba}, Y., {Utsumi}, Y., {et~al.} 2017, Publication of Korean
  Astronomical Society, 32, 225

\bibitem[{{Huang} {et~al.}(2017){Huang}, {Goto}, {Hashimoto}, {Oi}, \&
  {Matsuhara}}]{Huang17}
{Huang}, T.-C., {Goto}, T., {Hashimoto}, T., {Oi}, N., \& {Matsuhara}, H. 2017,
  \mnras, 471, 4239

\bibitem[{{Ilbert} {et~al.}(2006){Ilbert}, {Arnouts}, {McCracken},
  {Bolzonella}, {Bertin}, {Le F{\`e}vre}, {Mellier}, {Zamorani}, {Pell{\`o}},
  {Iovino}, {Tresse}, {Le Brun}, {Bottini}, {Garilli}, {Maccagni}, {Picat},
  {Scaramella}, {Scodeggio}, {Vettolani}, {Zanichelli}, {Adami}, {Bardelli},
  {Cappi}, {Charlot}, {Ciliegi}, {Contini}, {Cucciati}, {Foucaud}, {Franzetti},
  {Gavignaud}, {Guzzo}, {Marano}, {Marinoni}, {Mazure}, {Meneux}, {Merighi},
  {Paltani}, {Pollo}, {Pozzetti}, {Radovich}, {Zucca}, {Bondi}, {Bongiorno},
  {Busarello}, {de La Torre}, {Gregorini}, {Lamareille}, {Mathez}, {Merluzzi},
  {Ripepi}, {Rizzo}, \& {Vergani}}]{Ilbert06}
{Ilbert}, O., {Arnouts}, S., {McCracken}, H.~J., {et~al.} 2006, \aap, 457, 841

\bibitem[{{Jarrett} {et~al.}(2011){Jarrett}, {Cohen}, {Masci}, {Wright},
  {Stern}, {Benford}, {Blain}, {Carey}, {Cutri}, {Eisenhardt}, {Lonsdale},
  {Mainzer}, {Marsh}, {Padgett}, {Petty}, {Ressler}, {Skrutskie}, {Stanford},
  {Surace}, {Tsai}, {Wheelock}, \& {Yan}}]{Jarrett11}
{Jarrett}, T.~H., {Cohen}, M., {Masci}, F., {et~al.} 2011, \apj, 735, 112

\bibitem[{{Kartaltepe} {et~al.}(2010){Kartaltepe}, {Sanders}, {Le Floc'h},
  {Frayer}, {Aussel}, {Arnouts}, {Ilbert}, {Salvato}, {Scoville}, {Surace},
  {Yan}, {Brusa}, {Capak}, {Caputi}, {Carollo}, {Civano}, {Elvis}, {Faure},
  {Hasinger}, {Koekemoer}, {Lee}, {Lilly}, {Liu}, {McCracken}, {Schinnerer},
  {Smol{\v c}i{\'c}}, {Taniguchi}, {Thompson}, \& {Trump}}]{Kartaltepe10}
{Kartaltepe}, J.~S., {Sanders}, D.~B., {Le Floc'h}, E., {et~al.} 2010, \apj,
  709, 572

\bibitem[{{Kauffmann} \& {Haehnelt}(2000)}]{Kauffmann00}
{Kauffmann}, G., \& {Haehnelt}, M. 2000, \mnras, 311, 576

\bibitem[{{Kennicutt} {et~al.}(1987){Kennicutt}, {Keel}, {van der Hulst},
  {Hummel}, \& {Roettiger}}]{Kennicutt87}
{Kennicutt}, Jr., R.~C., {Keel}, W.~C., {van der Hulst}, J.~M., {Hummel}, E.,
  \& {Roettiger}, K.~A. 1987, \aj, 93, 1011

\bibitem[{{Kim} {et~al.}(2012){Kim}, {Lee}, {Matsuhara}, {Wada}, {Oyabu}, {Im},
  {Jeon}, {Kang}, {Ko}, {Lee}, {Takagi}, {Pearson}, {White}, {Jeong},
  {Serjeant}, {Nakagawa}, {Ohyama}, {Goto}, {Takeuchi}, {Pollo}, {Solarz}, \&
  {P{\c e}piak}}]{Kim12}
{Kim}, S.~J., {Lee}, H.~M., {Matsuhara}, H., {et~al.} 2012, \aap, 548, A29

\bibitem[{{Kim} {et~al.}(2015){Kim}, {Lee}, {Jeong}, {Goto}, {Matsuhara}, {Im},
  {Shim}, {Kim}, \& {Lee}}]{Kim15}
{Kim}, S.~J., {Lee}, H.~M., {Jeong}, W.-S., {et~al.} 2015, \mnras, 454, 1573

\bibitem[{{Lacy} {et~al.}(2007){Lacy}, {Petric}, {Sajina}, {Canalizo},
  {Storrie-Lombardi}, {Armus}, {Fadda}, \& {Marleau}}]{Lacy07}
{Lacy}, M., {Petric}, A.~O., {Sajina}, A., {et~al.} 2007, \aj, 133, 186

\bibitem[{{Lacy} {et~al.}(2015){Lacy}, {Ridgway}, {Sajina}, {Petric}, {Gates},
  {Urrutia}, \& {Storrie-Lombardi}}]{Lacy15}
{Lacy}, M., {Ridgway}, S.~E., {Sajina}, A., {et~al.} 2015, \apj, 802, 102

\bibitem[{{Lacy} {et~al.}(2004){Lacy}, {Storrie-Lombardi}, {Sajina},
  {Appleton}, {Armus}, {Chapman}, {Choi}, {Fadda}, {Fang}, {Frayer},
  {Heinrichsen}, {Helou}, {Im}, {Marleau}, {Masci}, {Shupe}, {Soifer},
  {Surace}, {Teplitz}, {Wilson}, \& {Yan}}]{Lacy04}
{Lacy}, M., {Storrie-Lombardi}, L.~J., {Sajina}, A., {et~al.} 2004, \apjs, 154,
  166

\bibitem[{{Lacy} {et~al.}(2013){Lacy}, {Ridgway}, {Gates}, {Nielsen}, {Petric},
  {Sajina}, {Urrutia}, {Cox Drews}, {Harrison}, {Seymour}, \&
  {Storrie-Lombardi}}]{Lacy13}
{Lacy}, M., {Ridgway}, S.~E., {Gates}, E.~L., {et~al.} 2013, \apjs, 208, 24

\bibitem[{{Lagache} {et~al.}(2003){Lagache}, {Dole}, \& {Puget}}]{Lagache03}
{Lagache}, G., {Dole}, H., \& {Puget}, J.-L. 2003, \mnras, 338, 555

\bibitem[{{Laurent} {et~al.}(2000){Laurent}, {Mirabel}, {Charmandaris},
  {Gallais}, {Madden}, {Sauvage}, {Vigroux}, \& {Cesarsky}}]{Laurent00}
{Laurent}, O., {Mirabel}, I.~F., {Charmandaris}, V., {et~al.} 2000, \aap, 359,
  887

\bibitem[{{Liu} \& {Kennicutt}(1995)}]{Liu95}
{Liu}, C.~T., \& {Kennicutt}, Jr., R.~C. 1995, \apj, 450, 547

\bibitem[{{Magdis} {et~al.}(2013){Magdis}, {Rigopoulou}, {Helou}, {Farrah},
  {Hurley}, {Alonso-Herrero}, {Bock}, {Burgarella}, {Chapman}, {Charmandaris},
  {Cooray}, {Dai}, {Dale}, {Elbaz}, {Feltre}, {Hatziminaoglou}, {Huang},
  {Morrison}, {Oliver}, {Page}, {Scott}, \& {Shi}}]{Magdis13}
{Magdis}, G.~E., {Rigopoulou}, D., {Helou}, G., {et~al.} 2013, \aap, 558, A136

\bibitem[{{Mateos} {et~al.}(2012){Mateos}, {Alonso-Herrero}, {Carrera},
  {Blain}, {Watson}, {Barcons}, {Braito}, {Severgnini}, {Donley}, \&
  {Stern}}]{Mateos12}
{Mateos}, S., {Alonso-Herrero}, A., {Carrera}, F.~J., {et~al.} 2012, \mnras,
  426, 3271

\bibitem[{{Matsuhara} {et~al.}(2006){Matsuhara}, {Wada}, {Matsuura},
  {Nakagawa}, {Kawada}, {Ohyama}, {Pearson}, {Oyabu}, {Takagi}, {Serjeant},
  {White}, {Hanami}, {Watarai}, {Takeuchi}, {Kodama}, {Arimoto}, {Okamura},
  {Lee}, {Pak}, {Im}, {Lee}, {Kim}, {Jeong}, {Imai}, {Fujishiro}, {Shirahata},
  {Suzuki}, {Ihara}, \& {Sakon}}]{Matsuhara06}
{Matsuhara}, H., {Wada}, T., {Matsuura}, S., {et~al.} 2006, \pasj, 58, 673

\bibitem[{{Miyazaki} {et~al.}(2012){Miyazaki}, {Komiyama}, {Nakaya}, {Kamata},
  {Doi}, {Hamana}, {Karoji}, {Furusawa}, {Kawanomoto}, {Morokuma}, {Ishizuka},
  {Nariai}, {Tanaka}, {Uraguchi}, {Utsumi}, {Obuchi}, {Okura}, {Oguri},
  {Takata}, {Tomono}, {Kurakami}, {Namikawa}, {Usuda}, {Yamanoi}, {Terai},
  {Uekiyo}, {Yamada}, {Koike}, {Aihara}, {Fujimori}, {Mineo}, {Miyatake},
  {Yasuda}, {Nishizawa}, {Saito}, {Tanaka}, {Uchida}, {Katayama}, {Wang},
  {Chen}, {Lupton}, {Loomis}, {Bickerton}, {Price}, {Gunn}, {Suzuki},
  {Miyazaki}, {Muramatsu}, {Yamamoto}, {Endo}, {Ezaki}, {Itoh}, {Miwa},
  {Yokota}, {Matsuda}, {Ebinuma}, \& {Takeshi}}]{Miyazaki12}
{Miyazaki}, S., {Komiyama}, Y., {Nakaya}, H., {et~al.} 2012, in \procspie, Vol.
  8446, Ground-based and Airborne Instrumentation for Astronomy IV, 84460Z

\bibitem[{{Murakami} {et~al.}(2007){Murakami}, {Baba}, {Barthel}, {Clements},
  {Cohen}, {Doi}, {Enya}, {Figueredo}, {Fujishiro}, {Fujiwara}, {Fujiwara},
  {Garcia-Lario}, {Goto}, {Hasegawa}, {Hibi}, {Hirao}, {Hiromoto}, {Hong},
  {Imai}, {Ishigaki}, {Ishiguro}, {Ishihara}, {Ita}, {Jeong}, {Jeong},
  {Kaneda}, {Kataza}, {Kawada}, {Kawai}, {Kawamura}, {Kessler}, {Kester},
  {Kii}, {Kim}, {Kim}, {Kobayashi}, {Koo}, {Kwon}, {Lee}, {Lorente}, {Makiuti},
  {Matsuhara}, {Matsumoto}, {Matsuo}, {Matsuura}, {M{\"U}ller}, {Murakami},
  {Nagata}, {Nakagawa}, {Naoi}, {Narita}, {Noda}, {Oh}, {Ohnishi}, {Ohyama},
  {Okada}, {Okuda}, {Oliver}, {Onaka}, {Ootsubo}, {Oyabu}, {Pak}, {Park},
  {Pearson}, {Rowan-Robinson}, {Saito}, {Sakon}, {Salama}, {Sato}, {Savage},
  {Serjeant}, {Shibai}, {Shirahata}, {Sohn}, {Suzuki}, {Takagi}, {Takahashi},
  {Tanab{\'E}}, {Takeuchi}, {Takita}, {Thomson}, {Uemizu}, {Ueno}, {Usui},
  {Verdugo}, {Wada}, {Wang}, {Watabe}, {Watarai}, {White}, {Yamamura},
  {Yamauchi}, \& {Yasuda}}]{Murakami07}
{Murakami}, H., {Baba}, H., {Barthel}, P., {et~al.} 2007, \pasj, 59, S369

\bibitem[{{Onaka} {et~al.}(2007){Onaka}, {Matsuhara}, {Wada}, {Fujishiro},
  {Fujiwara}, {Ishigaki}, {Ishihara}, {Ita}, {Kataza}, {Kim}, {Matsumoto},
  {Murakami}, {Ohyama}, {Oyabu}, {Sakon}, {Tanab{\'e}}, {Takagi}, {Uemizu},
  {Ueno}, {Usui}, {Watarai}, {Cohen}, {Enya}, {Ootsubo}, {Pearson}, {Takeyama},
  {Yamamuro}, \& {Ikeda}}]{Onaka07}
{Onaka}, T., {Matsuhara}, H., {Wada}, T., {et~al.} 2007, \pasj, 59, S401

\bibitem[{{Polletta} {et~al.}(2007){Polletta}, {Tajer}, {Maraschi},
  {Trinchieri}, {Lonsdale}, {Chiappetti}, {Andreon}, {Pierre}, {Le F{\`e}vre},
  {Zamorani}, {Maccagni}, {Garcet}, {Surdej}, {Franceschini}, {Alloin},
  {Shupe}, {Surace}, {Fang}, {Rowan-Robinson}, {Smith}, \&
  {Tresse}}]{Polletta07}
{Polletta}, M., {Tajer}, M., {Maraschi}, L., {et~al.} 2007, \apj, 663, 81

\bibitem[{{Puget} {et~al.}(1985){Puget}, {Leger}, \& {Boulanger}}]{Puget85}
{Puget}, J.~L., {Leger}, A., \& {Boulanger}, F. 1985, \aap, 142, L19

\bibitem[{{Sanders} {et~al.}(1988){Sanders}, {Soifer}, {Elias}, {Neugebauer},
  \& {Matthews}}]{Sanders88}
{Sanders}, D.~B., {Soifer}, B.~T., {Elias}, J.~H., {Neugebauer}, G., \&
  {Matthews}, K. 1988, \apjl, 328, L35

\bibitem[{{Stern} {et~al.}(2005){Stern}, {Eisenhardt}, {Gorjian}, {Kochanek},
  {Caldwell}, {Eisenstein}, {Brodwin}, {Brown}, {Cool}, {Dey}, {Green},
  {Jannuzi}, {Murray}, {Pahre}, \& {Willner}}]{Stern05}
{Stern}, D., {Eisenhardt}, P., {Gorjian}, V., {et~al.} 2005, \apj, 631, 163

\bibitem[{{Stern} {et~al.}(2012){Stern}, {Assef}, {Benford}, {Blain}, {Cutri},
  {Dey}, {Eisenhardt}, {Griffith}, {Jarrett}, {Lake}, {Masci}, {Petty},
  {Stanford}, {Tsai}, {Wright}, {Yan}, {Harrison}, \& {Madsen}}]{Stern12}
{Stern}, D., {Assef}, R.~J., {Benford}, D.~J., {et~al.} 2012, \apj, 753, 30

\bibitem[{{Weigel} {et~al.}(2018){Weigel}, {Schawinski}, {Treister},
  {Trakhtenbrot}, \& {Sanders}}]{Weigel18}
{Weigel}, A.~K., {Schawinski}, K., {Treister}, E., {Trakhtenbrot}, B., \&
  {Sanders}, D.~B. 2018, \mnras, 476, 2308

\bibitem[{{Yuan} {et~al.}(2010){Yuan}, {Kewley}, \& {Sanders}}]{Yuan10}
{Yuan}, T.-T., {Kewley}, L.~J., \& {Sanders}, D.~B. 2010, \apj, 709, 884

\end{thebibliography}
\end{document}